\documentclass[12pt,aps,showpacs]{revtex4}
\usepackage[dvips]{graphicx}
\usepackage{amssymb}
\usepackage{amsmath}
\usepackage{amsmath, amsthm, amssymb, mathrsfs}
\usepackage[utf8]{inputenc}

\newcommand{\be}{\begin{equation}}
\newcommand{\ee}{\end{equation}}
\newcommand{\bea}{\begin{eqnarray}} 
\newcommand{\eea}{\end{eqnarray}}
\newcommand{\bes}{\begin{subequations}}
\newcommand{\ees}{\end{subequations}}
\newcommand{\pa}{\partial}

\begin{document}
\title{Liouville-like solutions in dilaton gravity with Gauss-Bonnet modifications}

\author{D. Bazeia, R. Menezes and A. Yu. Petrov}

\affiliation{Departamento de F\'{\i}sica, Universidade Federal da Para\'{\i}ba, \\ 58051-970, Jo\~ao Pessoa, PB, Brazil}
\email{bazeia,rms,petrov@fisica.ufpb.br}  

\begin{abstract}
We investigate nonperturbative dilatonic solutions of the wide class of the modified gravity models including the Gauss-Bonnet terms with a general $F(G)$ Lagrangian. We show that presence of the Liouville-like solutions is a characteristic feature of these models.
\end{abstract}


\maketitle

The discovery that the Universe is presently undergoing accelerated expansion \cite{Riess} has inspired several researchers to search for physical models which could explain this phenomenon. Among different theories applied for this purpose, one of the most interesting suggestions, beside of use of the cosmological constant \cite{Lambda} and quintessence \cite{quint}, is the idea of a modified gravity according to which the common Hilbert-Einstein gravity action is extended by additive terms depending on some scalars constructed on the base of the metric tensor. In the mostly used form, such a modification implies in the theory called $F(R)$ gravity where the Lagrangian is a function of the scalar curvature (see e.g. \cite{NO}). The cosmological implication of such models were discussed, for instance, in \cite{frcos}, and in our recent works \cite{bcmp,bcmp1} some static solutions for such models were found.

However, the $F(R)$ modification is not the unique possibility to change the action of gravity without introducing extra fields. One of the popular suggestions for the modified gravity action, beside of the $F(R)$ models and of the Weyl gravity \cite{Sha}, is the Gauss-Bonnet gravity whose different classical aspects (including properties of some solutions of the equations of motion) were studied in a series of papers \cite{GBonne}. Recently, the Gauss-Bonnet gravity models were applied also for the brane studies \cite{GBonne0}, including holography aspects \cite{hol}, and solving the problem of cosmic acceleration, within which the modified Gauss-Bonnet gravity was treated as an alternative explication of dark energy \cite{GBonne1,fG1}. Also, the Gauss-Bonnet gravity models were applied to studies of higher-dimensional gravity \cite{5dGB} and of black holes \cite{BH}. Some observational estimations for Gauss-Bonnet gravity are given in \cite{Am,fG}.

Another direction of current interest is related to cosmology. Since we will be working with a single space coordinate, our investigation is somehow similar
to the study which generalizes the FRW scenario by including the Gauss-Bonnet contribution, since there one requires time evolution, and time is similar to a space coordinate, unless for its signature. Recent investigations on this issue appeared in \cite{GG}, and there one can also find the Liouville-like behavior
which will appear in the present study. The string-inspired investigations done in \cite{GG} and in references therein further stimulates the present study,
bridging its contents to cosmology and the up to now misterious dark energy fuel.

The interest to modified gravity models naturally stimulates the search the interest for the exact solutions in these models. In our papers \cite{bcmp,bcmp1} we have shown that the Liouville-like solutions, being the simplest examples of the exact solutions and, in certain cases, corresponding to the anti-de-Sitter space, are the typical one for the $F(R)$ gravity models. Therefore, the very natural question  is whether such solutions are possible in the more general modified gravity models, that is, in different versions of the Gauss-Bonnet-like models for the gravity in arbitrary space-time dimensions. We note that this interest is further supported by the fact that the time-dependent Liouville-like solutions were shown to arise also in the cosmological context \cite{GG}.

First of all, let us again, just as in \cite{bcmp}, restrict ourselves to the simplest conformally flat space
\cite{di}, with the dynamics is now concentrated in the conformal sector of the gravity, and the metric tensor $g_{\mu\nu}(x)$ is given by 
\be
g_{\mu\nu}(x)=e^{\sigma(x)}\eta_{\mu\nu}, 
\ee 
with $\sigma(x)$
standing for the conformal factor, the dilaton field. We use metric $(+,- , \ldots,-)$ and work with dimensionless fields and coordinates, for simplicity. 

In this case, one can find that the Christoffel symbols look like 
\be
\Gamma^{\beta}_{\mu\nu}=\frac{1}{2}\left(\delta^{\beta}_{\mu}\delta^{\lambda}_{\nu}+\delta^{\beta}_{\nu}\delta^{\lambda}_{\mu}-\eta^{\beta\lambda}\eta_{\mu\nu}\right)\pa_{\lambda}\sigma.  
\ee 
The Riemann curvature tensor is equal to 
\bea
R_{\mu\rho\alpha\beta}&=&\pa_{\alpha}\Gamma_{\mu,\rho\beta}-\pa_{\beta}\Gamma_{\mu,\rho\alpha}+
\Gamma_{\mu,\nu\alpha}\Gamma^{\nu}_{\rho\beta}-\Gamma_{\mu,\nu\beta}\Gamma^{\nu}_{\rho\alpha}=\nonumber\\
&=&e^{\sigma}[\frac{1}{2}(\pa_{\alpha}\pa_{\rho}\sigma\eta_{\mu\beta}-\pa_{\beta}\pa_{\rho}\sigma\eta_{\mu\alpha})+\frac{1}{2}(\pa_{\beta}\pa_{\mu}\sigma\eta_{\alpha\rho}-\pa_{\alpha}\pa_{\mu}\sigma\eta_{\beta\rho})+\nonumber\\&+&
\frac{1}{4}(\pa_{\rho}\pa_{\beta}\sigma \eta_{\alpha\mu}-\pa_{\rho}\pa_{\alpha}\sigma \eta_{\beta\mu})+
\frac{1}{4}(\pa_{\mu}\pa_{\alpha}\sigma \eta_{\beta\rho}-\pa_{\mu}\pa_{\beta}\sigma \eta_{\alpha\rho})+\nonumber\\&+&
\frac{1}{4}\pa_{\lambda}\sigma\pa^{\lambda}\sigma(\eta_{\alpha\rho}\eta_{\beta\mu}-\eta_{\beta\rho}\eta_{\alpha\mu})],
\eea 
and this implies that the Ricci tensor in $D$-dimensional space-time is
\bea
R_{\rho\alpha}=R^{\mu}_{\rho\alpha\mu}&=&\frac{1}{2}\Big[(D-2)\pa_{\rho}\pa_{\alpha}\sigma+\eta_{\rho\alpha}\Box\sigma\Big]-
\frac{D-2}{4}\Big[\pa_{\rho}\sigma\pa_{\alpha}\sigma-\eta_{\rho\alpha}\pa^{\mu}\sigma\pa_{\mu}\sigma\Big].
\eea 
Thus, the scalar curvature has the form 
\be
\label{curv}
R=g^{\alpha\beta}R_{\alpha\beta}=(D-1)e^{-\sigma}\Big[\Box\sigma+\left(\frac{D-2}{4}\right)\pa^{\alpha}\sigma\pa_{\alpha}\sigma\Big].
\ee
The useful contractions are also
\bea
R_{\mu\nu\lambda\rho}R^{\mu\nu\lambda\rho}&=&e^{-2\sigma}(D-2)\left[\pa_a\pa_b\sigma\pa^a\pa^b\sigma+(\Box\sigma)^2+
\frac{1}{8}(D-1)(\pa^a\sigma\pa_a\sigma)^2+\right.\nonumber\\&+&\left.
\pa^a\sigma\pa_a\sigma\Box\sigma-\pa_a\sigma\pa_b\sigma\pa^a\pa^b\sigma
\right],
\eea
and
\bea
R_{\mu\nu}R^{\mu\nu}&=&e^{-2\sigma}\left[\frac{(D-2)^2}{4}\pa_a\pa_b\sigma\pa^a\pa^b\sigma+(\frac{3D-4}{4})(\Box\sigma)^2+\right.\nonumber\\&+&\left.
\frac{1}{16}(D-1)(D-2)^2(\pa^a\sigma\pa_a\sigma)^2+\right.\nonumber\\&+&\left.
\frac{(D-2)(2D-3)}{4}\pa^a\sigma\pa_a\sigma\Box\sigma-\frac{(D-2)^2}{4}\pa_a\sigma\pa_b\sigma\pa^a\pa^b\sigma
\right]
\eea
The generic invariant action, which looks like
\bea
\label{generic}
S=\int d^D x \sqrt{|g|}\Big(AR^{2a}+B(R_{\mu\nu\lambda\rho}R^{\mu\nu\lambda\rho})^b+C(R_{\mu\nu}R^{\mu\nu})^c\Big),
\eea
in the dilatonic sector takes the following form
\bea
S&=&\int d^D xe^{(D/2-2)\sigma}\left(A(D-1)^{2a}e^{-2a\sigma}\Big[\Box\sigma+\left(\frac{D-2}{4}\right)\pa^{\alpha}\sigma\pa_{\alpha}\sigma\Big]^{2a}+\right.
\nonumber\\&+&Be^{-2b\sigma}\left[(D-2)\pa_a\pa_b\sigma\pa^a\pa^b\sigma+(\Box\sigma)^2+
\frac{1}{8}(D-1)(D-2)(\pa^a\sigma\pa_a\sigma)^2+\right.\nonumber\\&+&\left.
(D-2)\pa^a\sigma\pa_a\sigma\Box\sigma-(D-2)\pa_a\sigma\pa_b\sigma\pa^a\pa^b\sigma
\right]^{2b}+\nonumber\\&+&Ce^{-2c\sigma}\left[\frac{(D-2)^2}{4}\pa_a\pa_b\sigma\pa^a\pa^b\sigma+(\frac{3D}{4}-1)(\Box\sigma)^2+
\frac{1}{16}(D-1)(D-2)^2(\pa^a\sigma\pa_a\sigma)^2+\right.\nonumber\\&+&\left.\left.
\frac{(D-2)(2D-3)}{4}\pa^a\sigma\pa_a\sigma\Box\sigma-\frac{(D-2)^2}{4}\pa_a\sigma\pa_b\sigma\pa^a\pa^b\sigma
\right]^c
\right).
\eea
The important case of this generic action is the Gauss-Bonnet action depending on the Gauss-Bonnet invariant of the form
\bea
G=R^2-4R_{\mu\nu}R^{\mu\nu}+R_{\mu\nu\lambda\rho}R^{\mu\nu\lambda\rho},
\eea
with the explicit expression for it having the form
\bea
G&=&e^{-2\sigma}(D-2)(D-3)\left[(\Box\sigma)^2-\pa_a\pa_b\sigma\pa^a\pa^b\sigma+\frac{(D-1)(D-4)}{16}(\pa^a\sigma\pa_a\sigma)^2+\right.\nonumber\\&+&\left.\pa_a\sigma\pa_b\sigma\pa^a\pa^b\sigma+\frac{(D-3)}{2}\pa^a\sigma\pa_a\sigma\Box\sigma
\right].
\eea
In particular, in four dimensions, the Gauss-Bonnet action is a surface term:
\bea
S_{GB}=\int d^4x \sqrt{|g|}G=\int d^4x e^{2\sigma}G=\int d^4x \pa_a\left(2(\pa^a\sigma\Box\sigma-\pa^a\pa^b\sigma\pa_b\sigma)+\pa^a\sigma\pa^b\sigma\pa_b\sigma\right).
\eea 
In two and three dimensions, the Gauss-Bonnet invariant is an identical zero.

In this paper, we will concentrate on the domain wall case. To do this, we suggest that the dilaton $\sigma$ depends only on one spacial coordinate, say $z$. So, $(\Box\sigma)^2=(\sigma^{\prime\prime})^2$, $\pa_a\pa_b\sigma\pa^a\pa^b\sigma=(\sigma^{\prime\prime})^2$, $(\pa^a\sigma\pa_a\sigma)^2=(\sigma^{\prime})^4$, $\pa_a\sigma\pa_b\sigma\pa^a\pa^b\sigma=(\sigma^{\prime})^2\sigma^{\prime\prime}$,
$\pa^a\sigma\pa_a\sigma\Box\sigma=(\sigma^{\prime})^2\sigma^{\prime\prime}$.

First of all, we can show that the Liouville-like kink solutions are characteristic ones for a wide class of dilaton gravity models, with we suggest the restriction up to the domain wall dynamics. Indeed, in this case the scalar curvature is given by, 
\bea
\label{scalcurv}
R=-(D-1)e^{-\sigma} \left[\sigma^{\prime\prime}+ \frac{D-2}{4}\sigma^{\prime2}\right]
\eea
and the two other scalars look like
\bea
R_{\mu\nu\lambda\rho}R^{\mu\nu\lambda\rho}&=&(D-1)e^{-2\sigma}\left[(\sigma^{\prime\prime})^2+
\frac{1}{8}(D-2)(\sigma^{\prime})^4
\right],
\eea
and
\bea
R_{\mu\nu}R^{\mu\nu}&=&(D-1)e^{-2\sigma}\left[\frac{D}{4}(\sigma^{\prime\prime})^2+
\frac{1}{16}(D-2)^2(\sigma^{\prime})^4+
\frac{(D-2)}{4}\sigma^{\prime\prime}(\sigma^{\prime})^2
\right]
\eea
Let us consider the generic action (\ref{generic}). This action, in the dilatonic sector, by use of the expressions above, can be expressed as
\bea
S&=&\int d^Dx e^{\frac{D}{2}\sigma}\Big[Ae^{-2a\sigma}\left(\alpha_1(\sigma^{\prime\prime})^2+
\alpha_2(\sigma^{\prime})^4+\alpha_3\sigma^{\prime\prime}(\sigma^{\prime})^2\right)^a+\nonumber\\&+&Be^{-2b\sigma}\left(\beta_1(\sigma^{\prime\prime})^2+
\beta_2(\sigma^{\prime})^4+\beta_3\sigma^{\prime\prime}(\sigma^{\prime})^2\right)^b+\nonumber\\&+&Ce^{-2c\sigma}\left(\gamma_1(\sigma^{\prime\prime})^2+
\gamma_2(\sigma^{\prime})^4+\gamma_3\sigma^{\prime\prime}(\sigma^{\prime})^2\right)^c
\Big],
\eea
where $\alpha_i,\beta_i,\gamma_i,A,B,C$ are some constants whose exact form is not important.

We can obtain the equations of motion for this theory. We note that inside each parentheses, the number of derivatives in each term is equal to 4, thus, the number of derivatives in any term in the equations of motion proportional to $A$ is $4a$, to $B$ is $4b$, to $C$ is $4c$. Thus, we find the equations of motion in the form
\bea
&&Ae^{(-2a+\frac{D}{2})\sigma}\sum_{k_1,k_2,k_3}a_{k_1k_2k_3}\prod_{k_1+2k_2+3k_3=4a}(\sigma^{\prime})^{k_1}(\sigma^{\prime\prime})^{k_2}(\sigma^{\prime\prime\prime})^{k_3}+\nonumber\\&+&Be^{(-2b+\frac{D}{2})\sigma}\sum_{n_1,n_2,n_3}b_{n_1n_2n_3}\prod_{n_1+2n_2+3n_3=4b}(\sigma^{\prime})^{n_1}(\sigma^{\prime\prime})^{n_2}(\sigma^{\prime\prime\prime})^{n_3}+\nonumber\\&+&Ce^{(-2c+\frac{D}{2})\sigma}\sum_{m_1,m_2,m_3}c_{m_1m_2m_3}\prod_{m_1+2m_2+3m_3=4c}(\sigma^{\prime})^{m_1}(\sigma^{\prime\prime})^{m_2}(\sigma^{\prime\prime\prime})^{m_3}=0,
\eea
where $a_{k_1k_2k_3}$, $b_{n_1n_2n_3}$, $c_{m_1m_2m_3}$ are some numerical coefficients.
Then, we introduce the known Liouville-like ansatz
\bea
\label{ansatz}
\sigma^{\prime}=Ne^{l\sigma},
\eea
which implies in
\bea
\label{ans2}
\sigma^{\prime\prime}=N^2le^{2l\sigma}, \quad\, \sigma^{\prime\prime\prime}=2N^3l^2e^{3l\sigma}. 
\eea
Substituting these ansatzes, we see that in the term proportional to $A$, the field dependent multiplier factorizes giving the factor $e^{((4l-2)a+\frac{D}{2})\sigma}$, whereas in the terms proportional to $B$ and $C$ we get, respectively, $e^{((4l-2)b+\frac{D}{2})\sigma}$ and $e^{((4l-2)c+\frac{D}{2})\sigma}$. As a result, the equation of motion is reduced to the algebraic one
\bea
\label{alg}
&&Ae^{(4l-2)a\sigma}N^{4a}\sum_{k_1,k_2,k_3}a_{k_1k_2k_3}\prod_{k_1+2k_2+3k_3=4a}l^{k_2+2k_3}+\nonumber\\&+&Be^{(4l-2)b\sigma}N^{4b}\sum_{n_1,n_2,n_3}b_{n_1n_2n_3}\prod_{n_1+2n_2+3n_3=4b}l^{n_2+2n_3}+\nonumber\\&+&Ce^{(4l-2)c\sigma}N^{4c}\sum_{m_1,m_2,m_3}c_{m_1m_2m_3}\prod_{m_1+2m_2+3m_3=4c}l^{m_2+2m_3}=0.
\eea
The different solutions of this algebraic equation for the coefficients can be found. The consistent cases are $a=b=c$, or $l=\frac{1}{2}$. Also the consistent cases are when any two of the three coefficients $A,B,C$ are equal to zero, or one of them is equal to zero whereas the two others carry the same exponential factors, e.g., $A=0$ and $b=c$.

The case $l=1/2$ is of special importance. Indeed, in four dimensions the scalar curvature (\ref{scalcurv}) for this case looks like
\bea
R=-3N^2,
\eea
which is a negative constant. Thus, in four dimensions the case $l=1/2$ corresponds to anti-de Sitter space, with the coefficients of (\ref{alg}) are now related.

We can also focus on another model. Let us find and solve the equations of motion for the model with a bit different structure, that is, with the action initially introduced in \cite{fG1} (see also \cite{GBonne1,O1} for study of some aspects of this theory):
\bea
\label{difstr}
S=\int d^4x \sqrt{|g|}(R+f(G)).
\eea
Here we restrict ourselves by four dimensions. For simplicity, let us first, just as in \cite{bcmp}, consider the domain wall case. Thus, we get $G=e^{-2\sigma}((\sigma^{\prime})^3)^{\prime}$ and $R=-\frac{3}{2}e^{-\sigma}(2\sigma^{\prime\prime}+(\sigma^{\prime})^2)$
As a first attempt, we study the action 
\bea
S=\int d^4x \sqrt{|g|}(aR+bG^2)\simeq \int d^4x \left[9 b e^{-2\sigma}(\sigma^{\prime\prime})^2(\sigma^{\prime})^4-\frac{3}{2}a e^{\sigma}(2\sigma^{\prime\prime}+(\sigma^{\prime})^2)\right].
\eea
Here the sign $\simeq$ denotes that we restrict the coordinate dependence to the domain wall case. The corresponding equations of motion look like
\bea
&&6be^{-2\sigma}(\sigma^{\prime\prime})^2(\sigma^{\prime})^4-6b((\sigma^{\prime})^4\sigma^{\prime\prime}e^{-2\sigma})^{\prime\prime}+12b((\sigma^{\prime\prime})^2(\sigma^{\prime})^3e^{-2\sigma})^{\prime}+\nonumber\\&+&ae^{\sigma}\sigma^{\prime\prime}+\frac{1}{2}ae^{\sigma}(\sigma^{\prime})^2=0.
\eea
The most natural ansatz is (\ref{ansatz}) implying in (\ref{ans2}) and in
\bea
\label{ansatz1}
&&(\sigma^{\prime\prime})^2-\sigma^{\prime\prime}(\sigma^{\prime})^2=l(l-1)N^4e^{4l\sigma}, \quad
(e^{m\sigma})^{\prime}=Nme^{(m+l)\sigma},\nonumber\\ 
&&(e^{m\sigma})^{\prime\prime}=N^2m(m+l)e^{(m+2l)\sigma}.
\eea
Its substitution yields
\bea
12\,lbN^6e^{(6l-3)\sigma}[N(7l-2)(6l-2)-l(14l-3)]-aN^2\sigma\left(2l+1\right)=0.
\eea
We have some distinct possibilities: first, we can choose $b=0$, which yields $l=-1/2$, leading to usual gravitation; second, we can choose $a=0$, which yields 
$N=\frac{l(14l-3)}{(7l-2)(6l-2)}$; and third, we can choose $l=1/2$ with $a,b\neq 0$, so the exponential factor goes to 1, that is, we reproduce the adS solution, and we have the quintic algebraic equation, $3bN^4(3N-4)=2a$, which allows to find $N$ once $a$ and $b$ are given.

Our next step is the model of the form (see f.e. \cite{GBonne1}):
\bea
S=\int d^4x \sqrt{|g|}(a R^n+bG^m),
\eea
which generalizes the model (\ref{difstr}). In the domain wall case, it can be presented as
\bea
S=\int d^4x e^{2\sigma}\left(\tilde{a}[e^{-\sigma}(-2\sigma^{\prime\prime}-(\sigma^{\prime})^2)]^n+
\tilde{b}[e^{-2\sigma}(\sigma^{\prime})^2\sigma^{\prime\prime}]^m
\right),
\eea
with $\tilde{a}=\left(\frac{3}{2}\right)^na$, $\tilde{b}=3^mb$.
In this case, the equation of motion looks like
\bea
&&\tilde{a}\Bigg[
(2-n) e^{(2-n)\sigma}(-2\sigma^{\prime\prime}-(\sigma^{\prime})^2)^n+\nonumber\\&+&
2n\big[-e^{\sigma}(e^{-(n-1)\sigma}(-2\sigma^{\prime\prime}-(\sigma^{\prime})^2)^{n-1})^{\prime\prime}
 -\left(e^{-(n-1)\sigma}(-2\sigma^{\prime\prime}-(\sigma^{\prime})^2)^{n-1}\right)^{\prime}e^{\sigma}\sigma^{\prime}\big]
\Bigg]+\nonumber\\&+&
\tilde{b}(m-1)e^{(2-2m)\sigma}\left(\frac{1}{2}(\sigma^{\prime})^2\sigma^{\prime\prime}
\right)^{m-3}
\left\{-\frac{1}{4}\left((\sigma^{\prime})^2\sigma^{\prime\prime}
\right)^3+\frac{m}{2}(\sigma^{\prime})^3(\sigma^{\prime\prime})^2
\left[\sigma^{\prime}(\sigma^{\prime\prime})^2+\frac{1}{2}(\sigma^{\prime})^2\sigma^{\prime\prime\prime}
\right]-\right.\nonumber\\&-&\left.\frac{1}{4}(\sigma^{\prime})^4(\sigma^{\prime\prime})^2\left((\sigma^{\prime\prime})^2+\sigma^{\prime}\sigma^{\prime\prime\prime}
\right)+
\frac{m(m-2)}{2}(\sigma^{\prime})^2\left[\sigma^{\prime}(\sigma^{\prime\prime})^2+\frac{1}{2}(\sigma^{\prime})^2\sigma^{\prime\prime\prime}
\right]^2\right.\nonumber\\&+&\left.\frac{m}{4}(\sigma^{\prime})^4\sigma^{\prime\prime}
\left[(\sigma^{\prime\prime})^3+3\sigma^{\prime}\sigma^{\prime\prime}\sigma^{\prime\prime\prime}
+\frac{1}{2}(\sigma^{\prime})^2\sigma^{\prime\prime\prime\prime}
\right]+
\frac{1}{4}(\sigma^{\prime})^4(\sigma^{\prime\prime})^2[(\sigma^{\prime\prime})^2+\sigma^{\prime}\sigma^{\prime\prime\prime}]
\right\}=0.
\eea
The case $\tilde{a}=0$ we will consider in details further, the case $\tilde{b}=0$ can be read off from \cite{bcmp}, so we are interested only in the case when both $\tilde{a}$ and $\tilde{b}$ differ from zero. To solve this equation, we carry out the substitution used in (\ref{ansatz},\ref{ansatz1}). 
Also, we introduce the notations
\bea
\label{t}
T&=&\frac{D-4}{16}N^4+\frac{l}{2}N^3,\nonumber\\
\tilde{T}&=&\frac{D-4}{16}N^4+\frac{l}{4}N^3,
\eea
Thus we get:
\bea
&&\tilde{a}e^{(2-n+2ln)\sigma}N^{2n}\left[(2-n)(-2l-1)+2n(-2l-1)^{n-1}(2l-1)(n-1)(2ln-l-n+2)\right]\nonumber\\
&+&\tilde{b}e^{(4ml+2-2m)\sigma}T^{m-3}
\left[(2-2m)T-4lm(m-1)\tilde{T}-12m\tilde{T}+\right.\nonumber\\&+&\left.
\frac{m}{2}N^2l^2(m-1)[16m-32+20l^2]+(8m^2-5m)N^2l^2\right]=0.
\eea
It is easy to see, that if $n=2m$, any $l$ is possible. Otherwise, there exists an adS solution $l=\frac{1}{2}$, for which any $m$ and $n$ are possible. Other parameters can be found by solving the algebraic equations.

The natural continuation is a possible domain wall solution for the pure Gauss-Bonnet action in an arbitrary space-time dimension. In the domain wall case, the Gauss-Bonnet term is reduced to
\bea
G&=&e^{-2\sigma}(D-1)(D-2)(D-3)\left[\frac{(D-4)}{16}(\sigma^{\prime})^4+\frac{1}{2}(\sigma^{\prime})^2\sigma^{\prime\prime}
\right]
\eea 
Let us now consider the case $D>4$ (which is important from the extra dimensions viewpoint),
and study the model with the action 
\bea
S=\int d^Dx \sqrt{|g|}G^n=\int d^Dx e^{(\frac{D}{2}-2n)\sigma}\left[\frac{(D-4)}{16}(\sigma^{\prime})^4+\frac{1}{2}(\sigma^{\prime})^2\sigma^{\prime\prime}
\right]^n,
\eea
where we have omitted the irrelevant overall factor.
The equations of motion look like
\bea
&&\left[\frac{(D-4)}{16}(\sigma^{\prime})^4+\frac{1}{2}(\sigma^{\prime})^2\sigma^{\prime\prime}
\right]^{n-3}
\left\{(\frac{D}{2}-2n)\left[\frac{(D-4)}{16}(\sigma^{\prime})^4+\frac{1}{2}(\sigma^{\prime})^2\sigma^{\prime\prime}
\right]^3-\right.\nonumber\\&-&\left.
n(n-1)\left[\frac{(D-4)}{16}(\sigma^{\prime})^4+\frac{1}{2}(\sigma^{\prime})^2\sigma^{\prime\prime}
\right]\left[\frac{(D-4)}{4}(\sigma^{\prime})^3+\sigma^{\prime}\sigma^{\prime\prime}
\right]\times\right.\nonumber\\&\times&\left.\left[\frac{D-4}{4}(\sigma^{\prime})^3\sigma^{\prime\prime}+\sigma^{\prime}(\sigma^{\prime\prime})^2+\frac{1}{2}(\sigma^{\prime})^2\sigma^{\prime\prime\prime}
\right]-\right.\\&-&\left.
\frac{3}{4}(D-4)\left[\frac{(D-4)}{16}(\sigma^{\prime})^4+\frac{1}{2}(\sigma^{\prime})^2\sigma^{\prime\prime}\right]^2(\sigma^{\prime})^2\sigma^{\prime\prime}
+\right.\nonumber\\&+&\left.
\frac{n(n-1)(n-2)}{2}(\sigma^{\prime})^2\left[\frac{D-4}{4}(\sigma^{\prime})^3\sigma^{\prime\prime}+\sigma^{\prime}(\sigma^{\prime\prime})^2+\frac{1}{2}(\sigma^{\prime})^2\sigma^{\prime\prime\prime}
\right]^2\right.\nonumber\\&+&\left.\frac{n(n-1)}{2}(\sigma^{\prime})^2\left[\frac{(D-4)}{16}(\sigma^{\prime})^4+\frac{1}{2}(\sigma^{\prime})^2\sigma^{\prime\prime}
\right]\times\right.\nonumber\\&\times&\left.\left[\frac{3(D-4)}{4}(\sigma^{\prime})^2(\sigma^{\prime\prime})^2+\frac{D-4}{4}(\sigma^{\prime})^3\sigma^{\prime\prime\prime}+(\sigma^{\prime\prime})^3+3\sigma^{\prime}\sigma^{\prime\prime}\sigma^{\prime\prime\prime}
+\frac{1}{2}(\sigma^{\prime})^2\sigma^{\prime\prime\prime\prime}
\right]+\right.\nonumber\\&+&\left.2n(n-1)\left[\frac{(D-4)}{16}(\sigma^{\prime})^4+\frac{1}{2}(\sigma^{\prime})^2\sigma^{\prime\prime}
\right]\left[\frac{D-4}{4}(\sigma^{\prime})^3\sigma^{\prime\prime}+\sigma^{\prime}(\sigma^{\prime\prime})^2+\frac{1}{2}(\sigma^{\prime})^2\sigma^{\prime\prime\prime}
\right]\sigma^{\prime}\sigma^{\prime\prime}
\right\}=0.\nonumber
\eea
Again, we carry out the substitution given by (\ref{ansatz},\ref{ansatz1}). After introducing the notations (\ref{t}),
we find that this substitution gives the following condition: either $T=0$, or
\bea
&&(\frac{D}{2}-2n)T-4n\tilde{T}[l(n-1)\tilde{T}+3]+\frac{n}{2}N^2l^2(n-1)[16n-32+20l^2]+\nonumber\\&+&(8n^2-5n)N^2l^2=0.
\eea
From here, we can relate $C$ and $l$, thus, we again have the Liouville-like kink solutions.

Also, we can study the model with the action
\bea
S=-\int d^Dx \sqrt{|g|}(R-\alpha G),
\eea
whose explicit form in the domain wall case is
\bea
S&=&(D-1)\int d^Dx\Big[e^{(D/2-1)\sigma}[\sigma^{\prime\prime}+\frac{D-2}{2}(\sigma^{\prime})^2]+\nonumber\\&+&
\alpha (D-2)(D-3)e^{(D/2-2)\sigma}[\frac{D-4}{16}(\sigma^{\prime})^4+\frac{1}{2}(\sigma^{\prime})^2\sigma^{\prime\prime}] 
\Big].
\eea
For $D>4$ the corresponding equation of motion are
\bea
\sigma^{\prime\prime}+\frac{D-2}{2}(\sigma^{\prime})^2-\frac{\alpha}{16}(D-3)(D-4)[{5(D-4)}(\sigma^{\prime})^4+48(\sigma^{\prime})^2\sigma^{\prime\prime}]=0.
\eea
Implementing again the Liouville-like ansatz (\ref{ansatz}) we get the following form of this equation:
\bea
 N^2e^{2l\sigma}(D+2l-2)-\frac{\alpha}{16}(D-3)(D-4)e^{(4l-1)\sigma}N^4[5(D-4)+48l]=0.
\eea
The possible solutions are: first, $\alpha=0$ (that lead us to usual Einstein-Hilbert gravity), and second, the case $l=1/2$, whose solution is
\be
\sigma(x)=2 \ln \left(\frac{2}{|N_1|(x-x_0)} \right) 
\ee  
with
\be
N_1 = - \sqrt{\frac{16D}{\alpha (D-3)(D-4)}}
\ee
Thus, we again have found the Liouville-like solutions.

In summary, in this work we have studied a very wide class of the gravity models whose action depends not only on scalar curvature but also on the Gauss-Bonnet invariant and on other invariants constructed on the base of the Riemann curvature tensor, being considered both in four-dimensional space-time and in higher dimensions. As we have shown, all the models admit Liouville-like static kink solutions. Therefore, we can conclude that the presence of such solutions is common for different gravity models. Moreover, we succeeded to show that all these models in four-dimensional space-time allow the adS solution, and this seems to indicate that the systems admit supersymmetric extensions.

We have considered solutions of the gravitational field equations which depend on a single spacial coordinate. It is natural to expect that in cosmology, where the metrics (and therefore the dilaton) depends only on time, would display similar behavior implying in the appearance of Liouville-like solutions. Within this context, we notice that the time-dependent Liouville-like solutions were shown to arise in the gravity context earlier, for instance, in \cite{GG}, a fact that extends the interest of this work to the cosmological scenario.

A very natural continuation of the present study could be the coupling of the dilaton to extra scalar matter fields. Such an extension could be applied to  study more realistic scenarios, for example, in the braneworld and in the brane/cosmology contexts, giving thus a natural generalization of \cite{bcmp1}.

{\bf Acknowledgments.}
This work was partially supported by CNPq, and by PRONEX-CNPq-FAPESQ. The work by A. Yu. P. has been supported by CNPq-FAPESQ DCR program, CNPq project No. 350400/2005-9.

\end{document}